\documentstyle[11pt,aaspp4]{article}

\def\lsim{\lower.5ex\hbox{$\; \buildrel < \over \sim \;$}}
\def\gsim{\lower.5ex\hbox{$\; \buildrel > \over \sim \;$}}

\begin{document}

\title{Early Galactic Li, Be and B: Implications on Cosmic Ray Origin}

\author{R. Ramaty}
\affil{Laboratory for High Energy Astrophysics, Goddard Space Flight 
Center, Greenbelt, MD 20771; ramaty@gsfc.nasa.gov}

\and

\author{R. E. Lingenfelter}
\affil{Center for Astrophysics and Space Sciences, University of 
California San Diego, La Jolla, CA 92093-0424; rlingenfelter@ucsd.edu}

\begin{abstract} 

Be abundances of old, low metallicity halo stars have major 
implications on cosmic-ray origin, requiring  acceleration out of 
fresh supernova ejecta. The observed, essentially constant Be/Fe 
fixes the Be production per SNII, allowing the determination of the 
energy supplied to cosmic rays per SNII. The results rule out 
acceleration out of the metal-poor ISM, and favor Be production at 
all epochs of Galactic evolution by cosmic rays having the same 
spectrum and source composition as those at the current epoch. 
Individual supernova acceleration of its own nucleosynthetic 
products and the collective acceleration by SN shocks of 
ejecta-enriched matter in the interiors of superbubbles have been 
proposed for such origin. The supernova acceleration efficiency is 
about 2\% for the refractory metals and 10\% for all the cosmic 
rays. 

\end{abstract}

\section{Introduction}

It has been known for almost three decades that cosmic-ray 
interactions in the interstellar medium (hereafter ISM) have an 
important role in producing the Galactic inventories of the light 
elements Li, Be and B, hereafter LiBeB (Reeves, Fowler \& Hoyle 
1970; for reviews see also Reeves 1994 and Ramaty, Kozlovsky \& 
Lingenfelter 1998a). However, not all the LiBeB isotopes are 
cosmic-ray produced. About 10\% of the $^7$Li inventory results from 
nucleosynthesis in the Big Bang (e.g. Spite \& Spite 1993), and 
nucleosynthesis in a variety of Galactic objects, including core 
collapse supernovae (SNIIs, Woosley \& Weaver 1995), novae (Hernanz 
et al. 1996) and giant stars (Plez, Smith \& Lambert 1993; 
Wallerstein \& Morrell 1994), can produce a large fraction of the 
remaining 90\%. Core collapse supernovae also contribute to $^{11}$B 
production via $^{12}$C spallation by neutrinos during the explosion 
(Woosley et al. 1990; Woosley \& Weaver 1995). But essentially all 
of the $^6$Li, Be and $^{10}$B, and about 50\% of the $^{11}$B are 
cosmic-ray produced.

Starting about a decade ago, observations with ground based 
telescopes led to Be abundance determinations in old, low 
metallicity halo stars (see Vangioni-Flam et al. 1998 for a recent 
compilation of the data). These stars, with Fe-to-H abundance ratios 
as low as 10$^{-3}$ of the solar value, were born in the early 
Galaxy and still preserve fossil records of conditions that existed 
in that epoch of Galactic evolution. Thus, as the Be is entirely 
cosmic-ray produced, the early Galactic data extend the time scale 
of cosmic-ray research from the 10$^7$ year mean age of the current 
epoch cosmic rays to the more than 10 billion year age of the 
Galaxy. In particular, the fact that the average early Galactic ISM, 
unlike the ISM of the current epoch, was almost totally devoid of 
metals (C and heavier atoms), provides new clues on the nature of 
the source material out of which cosmic rays are accelerated. As we 
shall see, the early Galactic Be data strongly suggests that the 
cosmic rays are accelerated out of fresh supernova ejecta rather 
than the average ISM, because if they had been accelerated out of 
the ISM, the Be in the early Galaxy would have been highly 
underproduced. To allow the direct acceleration of supernova ejecta 
before they mix into the average ISM, two related cosmic-ray origin 
scenarios were developed recently. The first considers the 
individual supernova shock acceleration of its own nucleosynthetic 
products (Lingenfelter, Ramaty \& Kozlovsky 1998), while the second 
addresses the collective acceleration by successive SN shocks of 
ejecta-enriched matter in the interiors of superbubbles (Higdon, 
Lingenfelter \& Ramaty 1998). Freshly formed dust grains in 
supernovae play a central role in both models, as we shall see.

\begin{figure}[t]
  \begin{center}
    \leavevmode
\epsfxsize=10.cm
\epsfbox{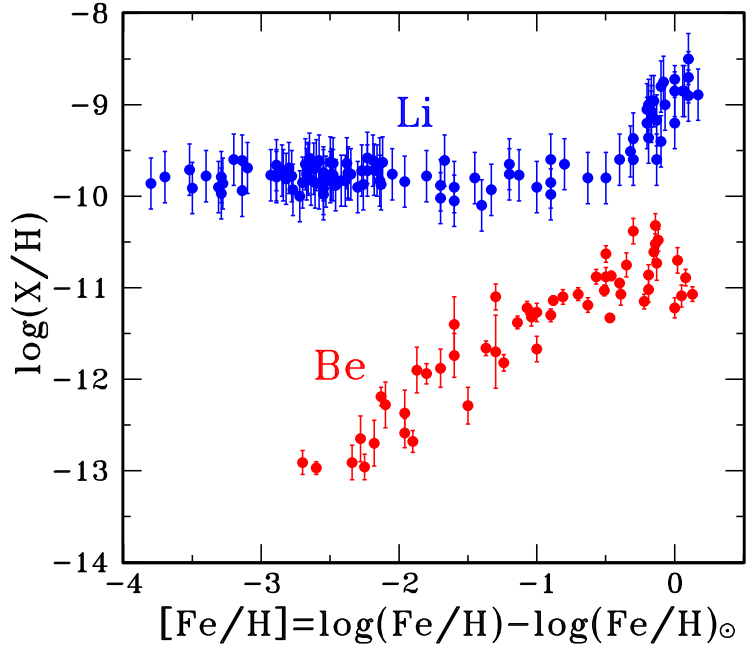}
  \end{center}
  \caption{Li and Be abundances for stars of various ages as a 
function of their Fe abundance. Data 
compilation by M. Lemoine for Li and Vangioni-Flam et al. (1998) 
for Be.} 
   \label{fig:1}
\end{figure} 

\section{LiBeB Origin}

Li and Be abundances for various stars as a function of their 
metallicity are shown in Fig.~1. The metallicity of a star is 
defined in terms of its Fe abundance, 
[Fe/H]$\equiv$log(Fe/H)$-$log(Fe/H)$_\odot$, where Fe/H is the Fe 
abundance by number relative to H and (Fe/H)$_\odot$ is the solar 
system value. [Fe/H] increases with time, reflecting the 
accumulating production of supernova nucleosynthesis, and thus 
provides a convenient, but nonlinear, representation of elapsed time 
since the formation of the Galaxy. Studies of Galactic chemical 
evolution (e.g. Ramaty, Lingenfelter \& Kozlovsky 1998b; see also 
Pagel 1997 and references therein) have provided information on the 
age-metallicity relation. The halo phase of our Galaxy, for which 
[Fe/H]$\lsim-$1, corresponds to a period of $\sim$10$^9$ years 
preceding the formation of the Galactic disk. LiBeB observations of 
the early Galaxy (e.g. Molaro et al. 1997; Hobbs \& Thorburn 1997; 
Duncan et al. 1997; Garcia Lopez et al. 1998) are very challenging 
because their spectral lines are weak and they are usually blended 
with interfering lines from other more abundant elements. The 
observations and their interpretation therefore require large 
telescopes and very efficient, high resolution detectors. While the 
Li and Be lines can be observed from the ground, the B lines require 
observations from space (with the Hubble Space Telescope, see Duncan 
et al. 1997). 

The flat portion of the Li evolution (Fig.~1), usually referred to 
as the Spite plateau (Spite \& Spite 1993), is generally believed 
(see Reeves 1994) to represent the $^7$Li abundance resulting from 
nucleosynthesis in the Big Bang. The subsequent increase in Li/H is 
due to nucleosynthesis in the various Galactic objects mentioned in 
the Introduction. Unlike Li/H, Be/H  increases with increasing 
[Fe/H] at all metallicities, implying a Galactic origin for Be even 
at the lowest metallicities where log(Be/H)$\simeq -13$. Indeed, the 
maximum contribution of Big Bang nucleosynthesis is insignificant, 
log(Be/H)$_{\rm BBNS}$$\simeq -15$ (Orito et al. 1997).

The implications of the Be data for cosmic-ray origin become more 
obvious (Ramaty et al. 1998a) when log(Be/Fe), rather than log(Be/H) 
is considered (Fig.~2). Here the horizontal line provides 
(Vangioni-Flam et al. 1998) the best fit to the data for 
[Fe/H]$<-1$. Fe production in this epoch (Timmes, Woosley \& Weaver 
1995) is dominated by SNIIs, which result from massive 
($>$10M$_\odot$), short lived ($\lsim$30 Myr) stellar progenitors 
with an IMF (initial mass function) averaged Fe yield per SNII of 
$\sim$0.1M$_\odot$, essentially independent of metallicity (Woosley 
\& Weaver 1995). The decrease of Be/Fe for [Fe/H]$\gsim -1$ most 
likely results from the additional Fe production in Type Ia 
supernovae (Timmes et al. 1995). These come from more slowly 
evolving white dwarf systems and are less frequent than the SNIIs, 
but because they do not produce neutron star remnants, they eject 
much more Fe per supernova and thus account for about half of the 
present Fe production.

\begin{figure}[t]
  \begin{center}
    \leavevmode
\epsfxsize=10.cm
\epsfbox{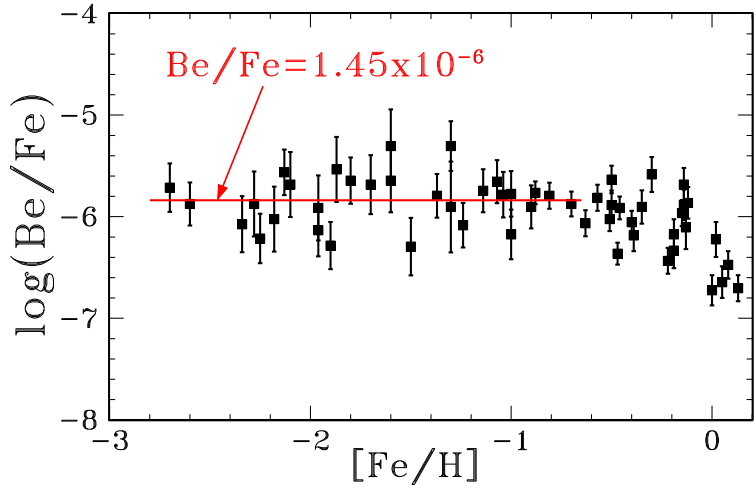}
  \end{center}
  \caption{Observed Be-Fe abundance ratio as a function 
of [Fe/H]; data compilation Vangioni-Flam et al. (1998). The  best fit, for 
[Fe/H]$\lsim$-1, implies that 2.3$\times$10$^{-8}$M$_\odot$ of Be 
are produced per average SNII. The decrease at higher [Fe/H] is due 
to contributions of Type Ia supernovae which make Fe but very 
little Be.} 
   \label{fig:2}
\end{figure}

The essentially constant observed Be-to-Fe abundance ratio for 
[Fe/H]$\lsim -1$, coupled with the fact that Fe in this epoch is 
produced only by SNIIs with a yield independent of [Fe/H], requires 
an essentially constant Be production per SNII,
\begin{eqnarray}
{\rm Q}_{\rm Be} \simeq 0.1\times 1.45\times 10^{-6}\times (9/56) = 
2.3\times10^{-8} {\rm M_\odot}~,
\end{eqnarray} 
independent of [Fe/H]. This is in conflict with cosmic-ray 
acceleration out of the ISM because in that case the composition of 
the cosmic rays (particularly C/H and O/H) would evolve in 
proportion to that of the ISM and Q$_{\rm Be}$ would increase with 
[Fe/H], contrary to the requirements of the data.

Independent evidence against cosmic ray acceleration purely out of 
the ISM is provided by energetics. The energy in cosmic rays per 
SNII, W$_{\rm SNII}$, needed to produce the required amount of Be 
depends on the composition of both the ISM and the cosmic rays, on 
the energy spectrum of the cosmic rays, and on the cosmic-ray escape 
length from the Galaxy, X$_{\rm esc}$ measured in g cm$^{-2}$ 
(Ramaty, Kozlovsky, Lingenfelter \& Reeves 1997). In Fig.~3 (from 
Ramaty et al. 1998b) we show W$_{\rm SNII}$, as a function of [Fe/H] 
for two values of X$_{\rm esc}$, and for two Galactic cosmic-ray 
origin models: a proposed CRS model for which the cosmic rays at all 
[Fe/H] have the same source composition and spectrum as the current 
epoch cosmic rays, and the ISM model for which the cosmic rays are 
accelerated out of a metallicity dependent interstellar medium with 
an energy spectrum that is also identical to that of the current 
epoch cosmic rays. The ambient ISM composition for both models is 
solar, scaled with 10$^{\rm [Fe/H]}$, except that O/H is allowed to 
increase by a factor of 3 for [Fe/H]$<-1$ to allow for the well 
known increase of O/H at low metallicities relative to its solar 
value (see Pagel 1997). To take into account recent shock 
acceleration results (Ellison, Drury \& Meyer 1997), the cosmic-ray 
C/H and O/H relative to the corresponding metallicity dependent ISM 
values are also increased by factors of 1.5 and 2. We see that the 
ISM model requires that W$_{\rm SN}$, the cosmic-ray energy 
per SNII, not only be metallicity dependent, which is unlikely, but 
also untenably large, exceeding the total available ejecta kinetic 
energy ($\sim$1.5$\times$10$^{51}$ erg, Woosley \& Weaver 1995) when 
[Fe/H]$<-2$. This reinforces the previous conclusion that cosmic 
rays accelerated out of the average, metal poor ISM cannot be 
responsible for the Be production in the early Galaxy. 

For the CRS model, on the other hand, W$_{\rm SNII}$ is essentially 
constant (Fig.~3), equal to the very reasonable value of 
$\sim$10$^{50}$ erg/SNII, practically the same as the energy 
supplied per supernova to the current epoch cosmic rays 
(Lingenfelter 1992). That these two energies are consistent, led to 
a different cosmic-ray paradigm, direct acceleration out of fresh 
supernova ejecta, at least for the refractory metals (Ramaty et al. 
1998a; Lingenfelter, Ramaty \& Kozlovsky 1998; Higdon, Lingenfelter 
\& Ramaty 1998). The constancy of Be/Fe is a straightforward 
consequence of such a model and clearly such a cosmic-ray origin 
provides the simplest explanation for the origin of Be throughout 
the entire evolutionary history of our Galaxy. Moreover, as we shall 
see (\S 3), acceleration of fresh ejecta, rather than average ISM 
material, is to be expected since the hot phase of the interstellar 
medium, where shock acceleration is most efficient (Axford 1981), is 
probably highly enriched in fresh gas and dust from the same 
supernovae whose shocks accelerate the cosmic rays (Higdon et al. 
1998). There is also another, more complex scenario for Be 
production by a possible, separate low energy cosmic-ray (LECR) 
component, also accelerated from fresh nucleosynthetic matter 
(Cass\'e, Lehoucq \& Vangioni-Flam 1995; Ramaty, Kozlovsky \& 
Lingenfelter 1996.) Such LECRs might allow the acceleration of the 
standard Galactic cosmic rays out of the ISM at all epochs of 
Galactic evolution, including the current one, but the nuclear 
gamma-ray lines that would provide the only evidence for their 
existence have not yet been seen. We consider these issues in \S 3, 
but before that we briefly discuss the B and $^6$Li data.

\begin{figure}[t]
  \begin{center}
    \leavevmode
\epsfxsize=10.cm
\epsfbox{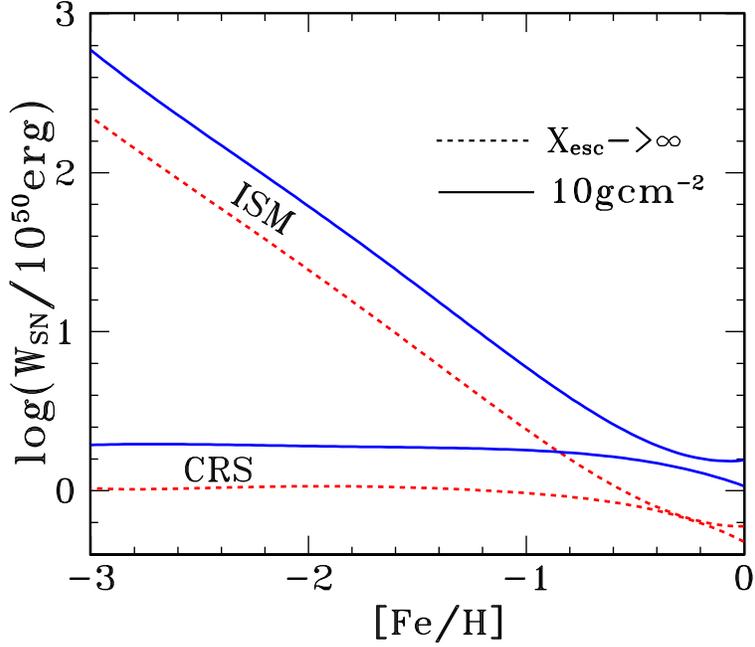}
  \end{center}
  \caption{Energy in cosmic rays per SNII required to produce 
2.3$\times$10$^{-8}$M$_\odot$ of Be. 
The cosmic ray source composition is metallicity 
independent for the CRS model and metallicity scaled for the ISM 
model. X$_{\rm esc}$$\simeq10$ g cm$^{-2}$ is the 
approximate current epoch 
cosmic-ray escape length; in the early Galaxy it could have been 
different, depending on the density and magnetic structure of the 
early Galactic halo. When X$_{\rm esc}$$\rightarrow$$\infty$, the 
cosmic rays are trapped in the halo until they are either stopped by 
Coulomb collisions or destroyed by nuclear reactions; this choice of 
escape length yields the lowest $W_{\rm SN}$ for the given  Be 
production.} 
   \label{fig:3}
\end{figure} 

Observations (Duncan et al. 1997; Garcia Lopez 1998) with the Hubble 
Space Telescope of the B abundance show (Fig.~4) that the B-to-Be 
abundance ratio is also essentially independent of [Fe/H], implying 
a common origin for these two elements. It has often been pointed 
out (e.g. Reeves 1994) that there is a problem with a pure 
cosmic-ray origin for B in that its isotopic ratio, 
$^{11}$B/$^{10}$B=4.05$\pm$0.2 measured in meteorites (Chaussidon \& 
Robert 1995) and $^{11}$B/$^{10}$B =3.4$\pm$0.7 in the interstellar 
medium (Lambert et al. 1998), exceeds the calculated ratio (2 to 
2.5) for production by the Galactic cosmic rays (Ramaty et al. 
1997). However, significant $^{11}$B production is also expected 
from $^{12}$C spallation by neutrinos in SNIIs (Woosley et al. 
1990). In fact, if $\sim$ 30\% of the $^{11}$B is from neutrinos the 
observed $^{11}$B/$^{10}$B can be explained, and since both the 
neutrino and cosmic-ray induced spallation processes are related to 
such supernovae, the constancy of the B-to-Be ratio is assured. The 
neutrinos mostly make $^{11}$B and not $^{10}$B because their 
temperature is not high enough for interactions above the higher 
threshold energy for $^{10}$B production. The required additional 
$^{11}$B production per SNII (Ramaty et al. 1997), about 
(2-7)$\times$10$^{-7}$ M$_\odot$, is consistent with the supernova 
calculations (Woosley \& Weaver 1995).

\begin{figure}[t]
  \begin{center}
    \leavevmode
\epsfxsize=10.cm
\epsfbox{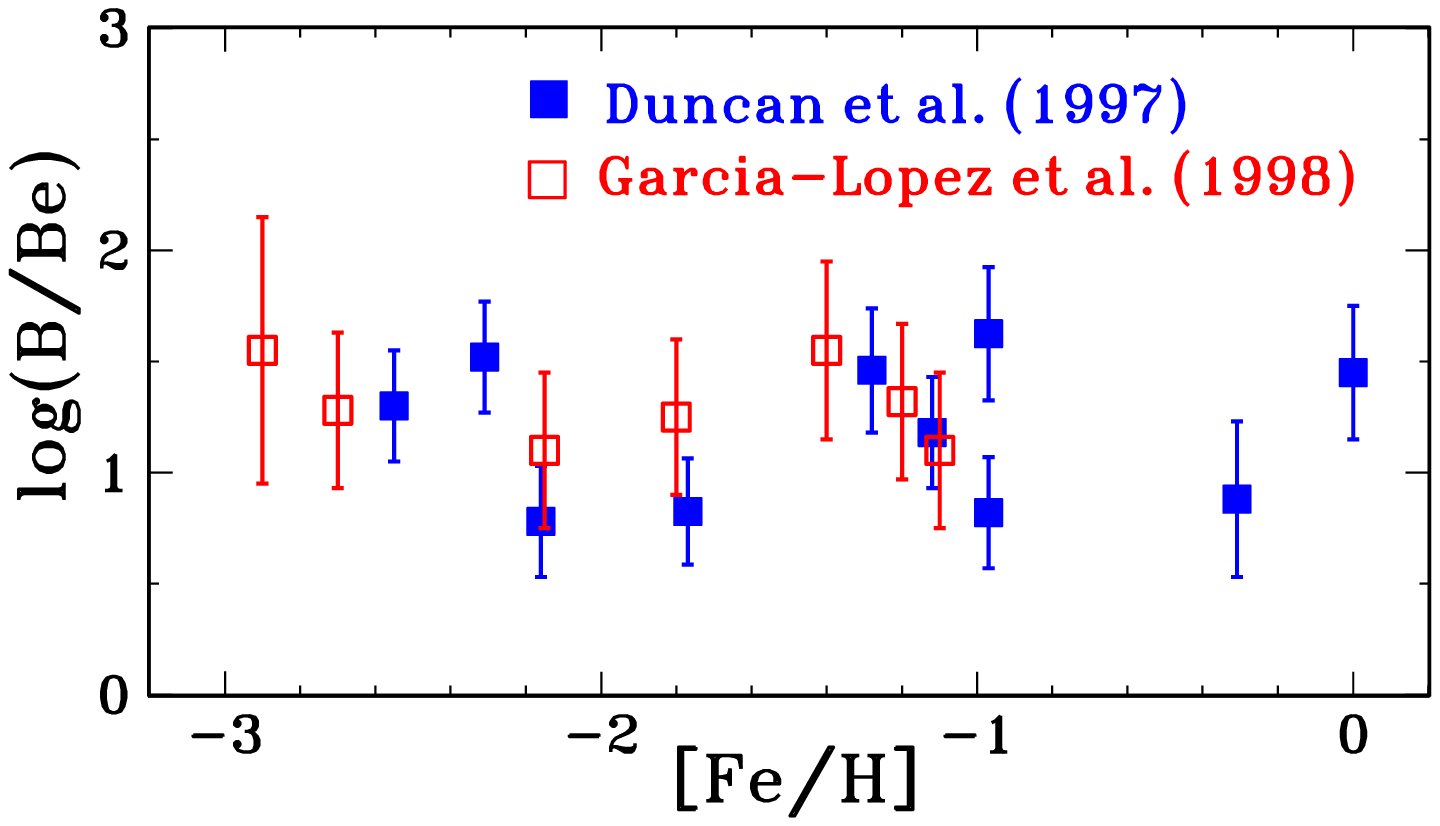}
  \end{center}
  \caption{Observations of the B-to-Be ratio as a function of 
[Fe/H]. The fact that this ratio does not vary much with metallicity 
implies a common origin for these two elements. SNII accelerated 
cosmic rays produce both B and Be and additional B comes from C 
spallation by neutrinos in SNIIs.} 
   \label{fig:4}
\end{figure}

The CRS model predicts (Ramaty et al. 1997) an abundance ratio 
$^6$Li/Be=5$\pm$0.5, essentially independent of [Fe/H] and 
consistent with the meteoritic value of 5.8 (Anders \& Grevesse 
1989). But at low metallicities ([Fe/H]$\simeq$-2.3), values of 
$^6$Li/Be as high as 60 are reported (Smith, Lambert \& Nissen 1998) 
which, even though still subject to large uncertainties, appear 
inconsistent with CRS production alone and would imply an additional 
source which dominated $^6$Li production at early times. Since 
significant $^6$Li can be produced by the $\alpha$-$\alpha$ reaction 
$^4$He($\alpha$,pn)$^6$Li whose cross section is very large below 
about 100 MeV/nucleon, $^6$Li/Be is strongly dependent on both the 
cosmic-ray composition (He/CNO) and energy spectrum, suggesting two 
possible additional sources. LECRs might account for these early 
Galactic $^6$Li data, but if $^6$Li/Be were to remain constant as a 
function of [Fe/H], as would be the case if the LECR component were 
also responsible for producing the Be at all metallicities, then 
$^6$Li/Be would be inconsistent with the meteoritic value (at 
[Fe/H]=0), and the total cosmic-ray produced Li abundance (which 
includes $^7$Li with $^7$Li/$^6$Li=1.5) would significantly exceed 
the Li/H data around [Fe/H]$\simeq -1$ (see Fig.~1). Alternatively, 
there could have been significant $^6$Li production via the 
$\alpha$-$\alpha$ reaction by possible pre-Galactic cosmic rays 
consisting almost entirely of primordial protons and $\alpha$ 
particles that would produce no Be. Further studies of this very 
exciting possibility, however, must await improvements in the 
reliability of the early Galactic $^6$Li measurements, which are 
very difficult. The contribution of Big Bang nucleosynthesis to 
Galactic $^6$Li again is very small (Nollett, Lemoine \& Schramm 
1997).   

\section{Cosmic-Ray Origin}

Motivated by the need to accelerate the cosmic rays out of fresh 
supernova ejecta, two related CRS scenarios were developed. The 
first considers the individual supernova shock acceleration of its 
own nucleosynthetic products (Lingenfelter et al. 1998), while the 
second addresses the collective acceleration by successive SN shocks 
of ejecta-enriched matter in the interiors of superbubbles (Higdon 
et al. 1998). In the individual supernova model, freshly formed high 
velocity grains in the slowing ejecta reach the forward supernova 
shock which then accelerates the grain erosion products. The 
superbubble model emphasizes the fact that the bulk of the SNIIs 
occur in the cores of supernova generated superbubbles where the 
ambient matter is likely to be dominated by fresh supernova ejecta. 
In both scenarios, grain erosion products play a central role. They 
provide an explanation for the observed cosmic-ray enrichment of the 
highly refractory Mg, Al, Si, Ca, Fe and Ni relative to the highly 
volatile H, He, N, Ne, Ar, an idea developed in detail previously 
for the ISM model (Meyer, Drury \& Ellison 1997). In both the 
individual supernova and superbubble scenarios, the accelerated C 
and O originate from grains, O from oxides (MgSiO$_3$, 
Fe$_3$O$_4$,Al$_2$O$_3$,CaO) and C mainly from graphite. As shown 
previously (Lingenfelter et al. 1998) such an origin for the C and O 
can explain the problematic C-to-O ratio in the cosmic rays which 
exceeds the corresponding solar ratio by about a factor of 2.

The similarity of the cosmic-ray source and solar abundance ratios 
of refractory elements, mainly Mg, Al, Si, Ca, relative to Fe, has 
been mentioned (Meyer et al. 1997) as an argument against the 
supernova ejecta origin for the cosmic rays. That this is not the 
case was demonstrated by Lingenfelter et al. (1998), the principal 
reason being that the combined contributions from SNIIs and Type Ia 
supernovae are responsible for both the cosmic-ray source and solar 
abundances of these refractories. 

The presence of s-elements in the cosmic rays has also been used as 
an argument against cosmic-ray acceleration out of supernova ejecta 
(e.g. Meyer et al. 1997), since such elements are not synthesized in 
the supernova explosions. However, this does not contradict 
acceleration out of the ejecta because s-elements are present in 
supernova ejecta along with the other much more abundant products of 
pre-supernova burning. The s-process elements are made in the cores 
of stars and can be ejected both in supernova explosions and in 
strong stellar winds (e.g. Arnould \& Takahashi 1993). In fact, the 
observations of SN1987A (e.g. Mazzali, Lucy \& Butler 1992 and 
references therein) show very significant overabundances of the 
prominent s-process products, Sr and Ba, relative to Fe by perhaps 
as much as an order of magnitude compared to solar values. Only a 
fraction of such supernova ejecta could account for the much more 
modest Sr and Ba overabundances of about 1.5, required for 
cosmic-ray source ratios (e.g. Binns et al. 1989). Moreover, the 
enrichment of r-process nuclei in the cosmic rays, especially the 
strong Pt peak (Waddington 1996), provides direct support for a 
supernova ejecta origin. The r-process elements are thought to be 
made just above the newly formed neutron star (e.g. Cardall \& 
Fuller 1997) in core collapse supernovae. Thus, both r- and 
s-process elements are blown off in the supernova ejecta along with 
the products of explosive burning and other products of earlier 
burning, and all the refractories condense in the ejecta.

Finally, electron-capture decay nuclei, such as $^{59}$Ni (decaying 
into $^{59}$Co with mean life of 1.1x10$^5$ yrs), can give a measure 
of the time between nucleosynthesis and acceleration, since such 
decay is suppressed once the nuclei are accelerated and fully 
ionized. Preliminary ACE data on the $^{59}$Ni-$^{59}$Co ratio 
(Wiedenbeck et al. 1998) show that $^{59}$Ni has decayed, 
suggesting delayed acceleration. This result favors the superbubble 
model for which the mean time between successive supernova 
explosions, $\sim$ 3$\times10^5$ yr, ensures that each supernova 
shock will on average accelerate ejecta accumulated from many 
previous supernovae on time scales clearly exceeding the $^{59}$Ni 
mean life. 

\section{Nuclear Gamma-Ray Line Emission}

The possible existence of a distinct low energy component of cosmic 
rays which could not be observed in the inner solar system because 
of solar modulation, is a topic of major interest for cosmic-ray 
research. Evidence for a strong enough LECR component that could 
produce significant amounts of LiBeB could only come from nuclear 
gamma-ray line data  (e.g. Ramaty, Kozlovsky \& Lingenfelter 1979). 
Indeed, as mentioned above, following the reported (Bloemen et al. 
1994) detection with COMPTEL/CGRO of C and O gamma-ray lines from 
the Orion star formation region, Cass\'e et al. (1995) suggested 
that the LECRs postulated to exist in the Orion region might be 
responsible for the Be (and B) production in the early Galaxy. The 
motivation for this idea was the indication (based on the reported 
spectrum of the line emission) that the LECRs in Orion are enriched 
in C and O relative to protons and $\alpha$ particles (see Ramaty 
1996 for review and Ramaty, Kozlovsky \& Lingenfelter 1996 for 
extensive calculations of LiBeB production by LECRs). It was 
proposed (Bykov 1995; Ramaty et al. 1996; Parizot, Cass\'e \& 
Vangioni-Flam 1997) that such enriched LECRs might be accelerated 
out of metal-rich winds of massive stars and the ejecta of 
supernovae from massive star progenitors ($>$60M$_\odot$) which 
explode within the bubble around the star formation region due to 
their very short lifetimes. These arguments led to the suggestion 
(Vangioni-Flam et al. 1996; Vangioni-Flam, Cass\'e \& Ramaty 1997) 
that the composition of the LECRs could be independent of Galactic 
metallicity, thus allowing them to reproduce the constancy of Be/Fe 
in the early Galaxy. The problem with this model is its energetic 
inefficiency, mostly because it relies on $>$60M$_\odot$ SNII 
progenitors which are much less numerous, but not significantly more 
energetic than the progenitors of all the SNIIs ($>$10M$_\odot$) 
which accelerate the cosmic rays in the CRS model. Moreover, the 
Orion gamma-ray line data have recently been withdrawn by the 
COMPTEL team (private communication, V. Sch\"onfelder, 1998). 
The planned high resolution observations of the Galaxy-wide nuclear 
gamma-ray line emission by the INTEGRAL mission should better define 
the possible contribution of LECRs to Galactic Be production.

\section{Conclusions}

We have seen how recent atomic spectroscopy observations of light 
element abundances in old halo stars have brought exciting new 
insights to the question of the origin of the cosmic rays, a problem 
that so far has been investigated mainly by in situ cosmic-ray 
observations. The cosmic rays in the early Galaxy, or at least their 
C and O, must have been accelerated from freshly nucleosynthesized 
matter rather than from the then extremely metal poor interstellar 
medium. This strongly suggests that the present epoch cosmic rays 
are also accelerated from fresh material, unlike in most current 
models. We have outlined two recently proposed scenarios for 
cosmic-ray acceleration from supernova ejecta that could account for 
such an origin.

\end{document}